\documentclass[12pt,thmsa]{article}
\usepackage{amssymb}

\input{tcilatex}
\begin{document}

\begin{titlepage}
\begin{center}
{\Large \bf Notation for general quantum teleportation}

\vspace{5mm}

\end{center}

\vspace{5 mm}

\begin{center}
{\bf Chong Li\footnote[1]{
lichong@student.dlut.edu.cn}, He-Shan Song and Yao-Xing
Luo\\}
\vspace{3mm}
{\it Department of Physics, Dalian University of Technology\\
Dalian, 116024, People's  Republic of China.}

\vspace{3mm}

\end{center}

\vspace{1cm}

\begin{center}
{\bf Abstract}
\end{center}
\baselineskip 24pt
A new notation for the quantum teleportation of finite dimensional quantum
state through a generally entangled quantum channel is introduced. For a
given quantum channel an explict mathematical criterion that governs the
faithful teleportation is presented.

\vspace{1cm}

\end{titlepage}
\baselineskip=24pt

Quantum teleportation has been introduced by Benett et al. \cite{bennett}
and discussed by a number of authors for two dimensional quantum state\cite
{l.v}\cite{s.l}\cite{s.bose}\cite{s.l.b}. By means of a classical and
distributed quantum communication channel, an unknown quantum state is
destroyed at the sending place while its perfect replica state, after some
rotation, appears at remote receiving place. Recently, for teleportation of
a general finite dimensional quantum state has been discussed in \cite{jp}
and \cite{Fei}, and the condition of the unitary transformation was given in
the\cite{Fei}.

In this letter, we introduced a new notation for quantum teleportation
through a generally entangled quantum channel and constructed an explicit
nitary transformation $U$ which receiver Bob performed on his qubit. It is
adapt to teleport a $n$-dimensional quantum state by an $\left( n+1\right) $%
-dimension generally entangled channel. At last, we given all the unitary
transformation $U$ for the case that both channel and measurement are in the 
\emph{Bell} state (maximally entangled state).

\vspace*{0.25cm}\vspace*{0.25cm}

Let \emph{H}$_1$ be the \emph{Hilbert} space with dimensions $N$, and $%
\left\{ \left| i\right\rangle _1\left| i=1...N\right. \right\} $ be the set
of orthogonal basis in \emph{H}$_1$. Alice has a general state $\left|
\varphi \right\rangle _1$ unknown to her in the \emph{Hilbert }space \emph{H}%
$_1$ as, 
\begin{equation}
\left| \varphi \right\rangle _1=\stackrel{N}{\stackunder{i=1}{\sum }}\alpha
_i\left| i\right\rangle _1.  \label{alice}
\end{equation}

Where $\alpha _i\in \mathfrak{C}$ ,and $\stackrel{N}{\stackunder{i=1}{\sum }}%
\left| \alpha _i\right| ^2=1$. For convenience we name an arbitrary
dimensional quantum state as a qubit.

Now we definite some useful notations$:$

\begin{definition}
$\mathbf{\alpha }=\left( 
\begin{array}{cccc}
\alpha _1 & \cdots & \alpha _{N-1} & \alpha _N
\end{array}
\right) $,
\end{definition}

\begin{definition}
$\mathbf{e}_i=\left( 
\begin{array}{cccc}
\left| 1\right\rangle _i & \cdots & \left| N-1\right\rangle _i & \left|
N\right\rangle _i
\end{array}
\right) $,
\end{definition}

\begin{definition}
$\mathbf{\tilde{e}}_{i=}\left( 
\begin{array}{cccc}
_i\left\langle 1\right| & \cdots & _i\left\langle N-1\right| & 
_i\left\langle N\right|
\end{array}
\right) $.
\end{definition}

Where the states $\left| 1\right\rangle _i\cdots \left| N\right\rangle _i$
represent the different states of the $i$-th qubit. For example, when $N=2$,
they can be spin-$\frac 12$ states, namely $\left| 1\right\rangle
\rightarrow \left| \uparrow \right\rangle $ and $\left| 2\right\rangle
\rightarrow \left| \downarrow \right\rangle $, and so on.

We rewrite the Eq.$\left( \ref{alice}\right) $ by the notation as 
\begin{equation}
\left| \varphi \right\rangle _1=\mathbf{\alpha e}_1^t\text{,}  \label{alice1}
\end{equation}

where $\mathbf{e}_i^t$ denotes transpose of $\mathbf{e}_i$.

Let \emph{H}$_2$ and \emph{H}$_3$ be auxiliary \emph{Hilbert} spaces
attached respectively to Alice and Bob, with dimensions $N_2=N_3=N$. A
generally entangled state of two qubits in the \emph{Hilbert} space \emph{H}$%
_2\otimes $\emph{H}$_3$ is given by 
\[
\left| \phi \right\rangle _{23}=\stackrel{N}{\stackunder{i\text{ }j}{\sum }}%
a_{ij}\left| i\right\rangle _2\left| j\right\rangle _3 
\]
with $_{23}\left\langle \phi \right. \left| \phi \right\rangle _{23}=1$.
Both $\left\{ \left| i\right\rangle _2\left| i=1...N\right. \right\} $ and $%
\left\{ \left| i\right\rangle _3\left| i=1...N\right. \right\} $ are sets of
orthogonal basis in \emph{Hilbert} spaces \emph{H}$_2$ and \emph{H}$_3$
respectively. We express $\left| \phi \right\rangle _{23}$ by the notation

\begin{equation}
\left| \phi \right\rangle _{23}=\mathbf{e}_2A\mathbf{e}_3^t\text{,}
\label{e23}
\end{equation}

where the known matrix $A=\left( 
\begin{array}{cccc}
a_{11} & a_{12} & \cdots & a_{1N} \\ 
a_{21} & a_{22} & \cdots & a_{2N} \\ 
\vdots & a_{i2} & \cdots & \vdots \\ 
a_{N1} & a_{N2} & \cdots & a_{NN}
\end{array}
\right) $for given quantum channel. The state for joint system of particle $%
1 $ and $2$ given to Alice and $3$ given to Bob is

\begin{equation}
\left| \Psi \right\rangle _{123}=\left| \varphi \right\rangle _1\otimes
\left| \phi \right\rangle _{23}=\mathbf{\alpha e}_1^t\mathbf{e}_2A\mathbf{e}%
_3^t=\mathbf{\alpha E}A\mathbf{e}_3^t  \label{123}
\end{equation}

with $\mathbf{E}=\mathbf{e}_1^t\mathbf{e}_2$.

To transform the state of Bob's qubit to be $\left| \varphi \right\rangle _3$%
,Bob need to do an unitary transformation on his qubit, after he obtain the
measured outcomes from Alice by classical channel. This process is called as
quantum teleportation.

Alice make a joint measurement on the first and the second qubits, then
particles $1$ and $2$ will form an entangled state $\left| \phi ^{\prime
}\right\rangle _{12}=\mathbf{e}_1B\mathbf{e}_2^t$, with $B=\left( 
\begin{array}{cccc}
b_{11} & b_{12} & \cdots & b_{1N} \\ 
b_{21} & b_{22} & \cdots & b_{2N} \\ 
\vdots & b_{i2} & \cdots & \vdots \\ 
b_{N1} & b_{N2} & \cdots & b_{NN}
\end{array}
\right) $. After Alice's measurement, Bob's particle $3$ will have been
projected into

\begin{equation}
\left| \varphi ^{^{\prime \prime }}\right\rangle _3=_{12}\left\langle \phi
^{^{\prime }}\right| \left. \Psi ^{}\right\rangle _{123}=\rho \mathbf{\alpha 
}X\mathbf{e}_3^t\text{,}  \label{b3}
\end{equation}

where $\left| \varphi ^{^{\prime \prime }}\right\rangle _{3}=\rho \left|
\varphi ^{^{\prime }}\right\rangle _{3}$, $\rho X=BA$, is a known matrix and 
$\rho $ is a known common factor,it is the probility amplitude of the state $%
\left| \varphi ^{^{\prime \prime }}\right\rangle _{3}$(if the channel and
measurement are \emph{Bell} state, $\rho =\frac{1}{2}$). Bob need to perform
a transformation $U$ on his qubit in order to obtain perfect replica of $%
\left| \varphi \right\rangle _{1}$ and hence realize the teleportation.

\begin{equation}
U\left| \varphi ^{^{\prime }}\right\rangle _3=\left| \varphi \right\rangle
_3=\mathbf{\alpha e}_3^t  \label{b3-1}
\end{equation}
{\Large Theorem}{\large \ \textbf{1} }If there is inverse matrix of the
known matrix $X^t$, the quantum teleportation can be realized and $\left(
X^t\right) ^{-1}$ is just the transformation $U$ that fulfills the quantum
teleportation.

{\emph{Proof : }}

By using the Eq.$\left( \ref{alice1}\right) $ $\left( \ref{e23}\right) $and$%
\left( \ref{123}\right) $, we have

\begin{eqnarray*}
\left| \varphi ^{^{\prime }}\right\rangle _3 &=&\stackrel{N}{\stackunder{i%
\text{ }j\text{ }k\text{ }m\text{ }n}{\sum }}\alpha
_ia_{jk}b_{mn2}\left\langle n\right| \left. j\right\rangle _{21}\left\langle
m\right| \left. i\right\rangle _1\left| k\right\rangle _3 \\
&=&\stackrel{N}{\stackunder{i\text{ }j\text{ }k\text{ }m\text{ }n}{\sum }}%
\alpha _ia_{jk}b_{mn}\delta _{mi}\delta _{nj}\left| k\right\rangle _3 \\
&=&\rho \stackrel{N}{\stackunder{i\text{ }k}{\sum }}\alpha _ix_{ik}\left|
k\right\rangle _3\left( \rho x_{ik}=\stackrel{N}{\stackunder{\text{ }j}{\sum 
}}a_{jk}b_{ij}\right)
\end{eqnarray*}

namely

\[
\left| \varphi ^{^{\prime \prime }}\right\rangle _3=\mathbf{\alpha }BA%
\mathbf{e}_3^t=\rho \mathbf{\alpha }X\mathbf{e}_3^t=\rho \left| \varphi
^{^{\prime }}\right\rangle _3\text{.} 
\]

Bob perform the operation $U$ on the state $\left| \varphi ^{^{\prime
}}\right\rangle _3$, then he will obtain perfect replica of $\left| \varphi
\right\rangle _1$,

\[
U\left| \varphi ^{^{\prime }}\right\rangle _3=\left| \varphi \right\rangle _3%
\text{,} 
\]

namely

\begin{equation}
U\stackrel{N}{\stackunder{\text{ }k}{\sum }}x_{ik}\left| k\right\rangle
_3=\left| i\right\rangle _3\text{.}  \label{*4}
\end{equation}

Let the basis $\left| i\right\rangle $ be an $N$-dimensional column vector

\begin{equation}
\left| i\right\rangle =\stackunder{1................i.................N}{%
\left( 
\begin{array}{ccccc}
0 & \cdots & 1 & \cdots & 0
\end{array}
\right) }^t\text{.}  \label{*5}
\end{equation}

From the Eq.$\left( \ref{*4}\right) $ and$\left( \ref{*5}\right) $, we have

\[
\stackrel{N}{\stackunder{\text{ }k}{\sum }}U_{ij}X_{kj}=\delta _{ik}\text{,} 
\]

so the matrix $U$ is nothing but the inverse matrix of the known matrix $X^t$

\[
UX^t=I\text{,} 
\]

\[
U=\left( X^t\right) ^{-1}\text{.}\blacksquare 
\]

From above discussion, we see that any finite state can be teleported
through a generally entangled state.

This theorem can be extended the teleportation of multi-qubit such as the
states discussed in the Ref.\cite{long}, if we re-define merely some
notation. We define the state to be teleportated 
\[
\left| \psi \right\rangle =\stackrel{N}{\stackunder{i=1}{\sum }}\alpha
_i\left| \stackunder{n}{\underbrace{ii\cdots i}}\right\rangle _1=\mathbf{%
\alpha e}_1^t\text{.} 
\]

with $\mathbf{e}_1=\left( \left| \stackunder{n}{\underbrace{1\cdots 1}}%
\right\rangle _1\left| \stackunder{n}{\underbrace{2\cdots 2}}\right\rangle
_1\cdots \left| \stackunder{n}{\underbrace{n\cdots n}}\right\rangle
_1\right) $,

the channel is 
\[
\left| \phi \right\rangle _{23}=\stackrel{N}{\stackunder{i=1}{\sum }}%
a_i\left| \stackunder{n+1}{\underbrace{ii\cdots i}}\right\rangle _{23}=%
\mathbf{e}_2A\mathbf{e}_3^t\text{,} 
\]

with $\mathbf{e}_2=\left( \left| 1\right\rangle _2\left| 2\right\rangle
_2\cdots \left| n\right\rangle _2\right) $ and $\mathbf{e}_3=\left( \left| 
\stackunder{n}{\underbrace{1\cdots 1}}\right\rangle _3\left| \stackunder{n}{%
\underbrace{2\cdots 2}}\right\rangle _3\cdots \left| \stackunder{n}{%
\underbrace{n\cdots n}}\right\rangle _3\right) $.

and Alice's measured result is

\[
\left| \phi ^{^{\prime }}\right\rangle _{12}=\mathbf{e}_1B\mathbf{e}_2\text{,%
} 
\]

then we can construct the transformation on the third qubit by using the
theorem

\[
U=\left( (BA)^t\right) ^{-1}/\rho \text{.} 
\]

It is easily show that the operation $U$ may be non-unitary. In this case,
the teleportation become probabilistic as the result given in Ref. \cite{gcg}%
.

In general, the Bob's transformation $U$ should be unitary, so a limitation
should be considered, namely the matrix $X\left( \text{or }X^t\right) $ must
satisfy the following equation,

\[
\left. 
\begin{array}{c}
\left( X^t\right) ^{\dagger }=\left( X^t\right) ^{-1} \\ 
\stackrel{N}{\stackunder{j=1}{\sum }}x_{ji}x_{kj}^{*}=\delta _{ik}
\end{array}
\right. 
\]

Therefore, if the channel and the Alice's measuremed result are known, and
the matrix $X$ is unitary, Bob can obtain the transported state. Neither the
channel or the measurement result must be maximally entangled state, or the
channel can be a maximally entangled state, but the Alice's measured result
can be non-maximum entangled state. For example, we take $N=2$, let the
quantum channel be a generally entangled state \vspace*{0.25cm}$\left| \psi
\right\rangle _{23}=\mathbf{e}_2A\mathbf{e}_3^t$, with $A=\left( 
\begin{array}{cc}
a & b \\ 
c & d
\end{array}
\right) $ and $\det \left( A\right) \neq 0$, and let the measured result of
Alice be $\left| \phi ^{^{\prime }}\right\rangle _{12}=\mathbf{e}_1B\mathbf{e%
}_2^t$, with $B=\left( 
\begin{array}{cc}
a^{^{\prime }} & b^{^{\prime }} \\ 
c^{^{\prime }} & d^{^{\prime }}
\end{array}
\right) $ and $\det \left( B\right) \neq 0$, then the unitary matrix $\left(
X\right) ^t=\left( BA\right) ^t/\rho =\left( 
\begin{array}{cc}
a^{^{\prime }}a+b^{^{\prime }}c & ac^{^{\prime }}+cd^{^{\prime }} \\ 
a^{^{\prime }}b+b^{^{\prime }}d & c^{^{\prime }}b+d^{^{\prime }}b
\end{array}
\right) \dfrac 1\rho =\left( 
\begin{array}{cc}
x_1 & x_3 \\ 
x_2 & x_4
\end{array}
\right) $. We can calculate out the unitary matrix $X$,

\begin{eqnarray*}
X &=&\left( 
\begin{array}{cc}
\cos \left( \theta \right) & \sin \left( \theta \right) \\ 
\sin \left( \theta \right) & -\cos \left( \theta \right)
\end{array}
\right) \\
\text{or }X &=&\left( 
\begin{array}{cc}
\cos \left( \theta \right) & -\sin \left( \theta \right) \\ 
\sin \left( \theta \right) & \cos \left( \theta \right)
\end{array}
\right) \text{,}
\end{eqnarray*}

so we can get the inverse matrix of $X^t$ as

\[
\left( X^t\right) ^{-1}=\left( X^t\right) ^{\dagger }\text{.} 
\]

If we assume that Alice's measurement is \emph{Bell} basis, namely $B\in
\left\{ \sigma _0\text{,}\sigma _1\text{,}\sigma _2\text{,}\sigma _3\right\} 
$, then the matrix $\left( X^t\right) ^{-1}$ can be obtained easily. We
given all the matrix $U$ in table $\left( 1\right) $ for the different
channels and different measured result, and all the channels and the
measured results are \emph{Bell} state.( all the matrixes have a common
factor $\rho =\tfrac 1{\sqrt{2}}$)

In addition, the theorem can also be used to the general case, as $\dim
\left( H_1\right) =n\leqslant \dim \left( H_2\right) =m$. Let $\mathbf{%
\alpha }$ and $\mathbf{e}_1$ be respectively

$\mathbf{\alpha }=\left( 
\begin{array}{llllll}
\alpha _1 & \cdots & \alpha _n & \alpha _{n+1} & \cdots & \alpha _m
\end{array}
\right) $ where $\alpha _i=0$ for $i=n+1$ $\cdots $ $m$

and $\mathbf{e}_1=\left( 
\begin{array}{llllll}
\left| 1\right\rangle _1 & \cdots & \left| n\right\rangle _1 & \left|
n+1\right\rangle _1 & \cdots & \left| m\right\rangle _1
\end{array}
\right) $.

If we can construct a matrix $U$, which made matrix $X^tU$ be diagonal and
all diagonal elements of matrix $X^tU$ be $0$ or $1$, the teleportation can
be realized. We can delete all of the elements from the $\left( n+1\right) $%
-th raw to the $m$-th raw and $\left( n+1\right) $-th column to the $m$-th
column of the matrix $X^t$ then we got matrix $X^{t^{\prime }}$, and
construct the inverse matrix of $X^{t^{\prime }}$ by the theorem $1$ easily.
Let $U=U^{^{\prime }}\oplus \mathbf{0}$ $\left( \text{with }U^{^{\prime
}}=\left( X^{t^{\prime }}\right) ^{-1}\right) $, then the matrix $U$ is just
the operation which Bob's performed on his qubit.

After introduced a new notation for the quantum teleportation, we obtained
the condition on quantum teleportation by a generally entangled channel, the
channel must not be maximally entangled state. We constructed the operation $%
U$ performed by the Bob on the third qubit in order to obtain perfect
replica of teleported state. We given the general form of the operator $U$
for the case that both channel and the state to be teleported are
two-dimensional.

\vspace{3cm}

\begin{tabular}{||l||l||l||l||l||}
\hline\hline
$A$ 1 & $B$ & $BA$ & $\left( BA\right) ^{t}$ & $U$ \\ \hline\hline
$\left( 
\begin{array}{cc}
1 & 0 \\ 
0 & 1
\end{array}
\right) $ & $\left( 
\begin{array}{cc}
1 & 0 \\ 
0 & 1
\end{array}
\right) $ & $\left( 
\begin{array}{cc}
1 & 0 \\ 
0 & 1
\end{array}
\right) $ & $\left( 
\begin{array}{cc}
1 & 0 \\ 
0 & 1
\end{array}
\right) $ & $\left( 
\begin{array}{cc}
1 & 0 \\ 
0 & 1
\end{array}
\right) $ \\ \hline\hline
$\left( 
\begin{array}{cc}
1 & 0 \\ 
0 & 1
\end{array}
\right) $ & $\left( 
\begin{array}{cc}
1 & 0 \\ 
0 & -1
\end{array}
\right) $ & $\left( 
\begin{array}{cc}
1 & 0 \\ 
0 & -1
\end{array}
\right) $ & $\left( 
\begin{array}{cc}
1 & 0 \\ 
0 & -1
\end{array}
\right) $ & $\left( 
\begin{array}{cc}
1 & 0 \\ 
0 & -1
\end{array}
\right) $ \\ \hline\hline
$\left( 
\begin{array}{cc}
1 & 0 \\ 
0 & 1
\end{array}
\right) $ & $\left( 
\begin{array}{cc}
0 & 1 \\ 
1 & 0
\end{array}
\right) $ & $\left( 
\begin{array}{cc}
0 & 1 \\ 
1 & 0
\end{array}
\right) $ & $\left( 
\begin{array}{cc}
0 & 1 \\ 
1 & 0
\end{array}
\right) $ & $\left( 
\begin{array}{cc}
0 & 1 \\ 
1 & 0
\end{array}
\right) $ \\ \hline\hline
$\left( 
\begin{array}{cc}
1 & 0 \\ 
0 & 1
\end{array}
\right) $ & $\left( 
\begin{array}{cc}
0 & -1 \\ 
1 & 0
\end{array}
\right) $ & $\left( 
\begin{array}{cc}
0 & -1 \\ 
1 & 0
\end{array}
\right) $ & $\left( 
\begin{array}{cc}
0 & 1 \\ 
-1 & 0
\end{array}
\right) $ & $\left( 
\begin{array}{cc}
0 & -1 \\ 
1 & 0
\end{array}
\right) $ \\ \hline\hline
\end{tabular}

\medskip

\begin{tabular}{||l||l||l||l||l||}
\hline\hline
$A$ 2 & $B$ & $BA$ & $\left( BA\right) ^{t}$ & $U$ \\ \hline\hline
$\left( 
\begin{array}{cc}
0 & 1 \\ 
1 & 0
\end{array}
\right) $ & $\left( 
\begin{array}{cc}
1 & 0 \\ 
0 & 1
\end{array}
\right) $ & $\left( 
\begin{array}{cc}
0 & 1 \\ 
1 & 0
\end{array}
\right) $ & $\left( 
\begin{array}{cc}
0 & 1 \\ 
1 & 0
\end{array}
\right) $ & $\left( 
\begin{array}{cc}
0 & 1 \\ 
1 & 0
\end{array}
\right) $ \\ \hline\hline
$\left( 
\begin{array}{cc}
0 & 1 \\ 
1 & 0
\end{array}
\right) $ & $\left( 
\begin{array}{cc}
1 & 0 \\ 
0 & -1
\end{array}
\right) $ & $\left( 
\begin{array}{cc}
0 & -1 \\ 
1 & 0
\end{array}
\right) $ & $\left( 
\begin{array}{cc}
0 & 1 \\ 
-1 & 0
\end{array}
\right) $ & $\left( 
\begin{array}{cc}
0 & -1 \\ 
1 & 0
\end{array}
\right) $ \\ \hline\hline
$\left( 
\begin{array}{cc}
0 & 1 \\ 
1 & 0
\end{array}
\right) $ & $\left( 
\begin{array}{cc}
0 & 1 \\ 
1 & 0
\end{array}
\right) $ & $\left( 
\begin{array}{cc}
1 & 0 \\ 
0 & 1
\end{array}
\right) $ & $\left( 
\begin{array}{cc}
1 & 0 \\ 
0 & 1
\end{array}
\right) $ & $\left( 
\begin{array}{cc}
1 & 0 \\ 
0 & 1
\end{array}
\right) $ \\ \hline\hline
$\left( 
\begin{array}{cc}
0 & 1 \\ 
1 & 0
\end{array}
\right) $ & $\left( 
\begin{array}{cc}
0 & -1 \\ 
1 & 0
\end{array}
\right) $ & $\left( 
\begin{array}{cc}
1 & 0 \\ 
0 & -1
\end{array}
\right) $ & $\left( 
\begin{array}{cc}
1 & 0 \\ 
0 & -1
\end{array}
\right) $ & $\left( 
\begin{array}{cc}
1 & 0 \\ 
0 & -1
\end{array}
\right) $ \\ \hline\hline
\end{tabular}

\medskip

\begin{tabular}{||l||l||l||l||l||}
\hline\hline
$A$ 3 & $B$ & $BA$ & $\left( BA\right) ^{t}$ & $U$ \\ \hline\hline
$\left( 
\begin{array}{cc}
0 & -1 \\ 
1 & 0
\end{array}
\right) $ & $\left( 
\begin{array}{cc}
1 & 0 \\ 
0 & 1
\end{array}
\right) $ & $\left( 
\begin{array}{cc}
0 & -1 \\ 
1 & 0
\end{array}
\right) $ & $\left( 
\begin{array}{cc}
0 & 1 \\ 
-1 & 0
\end{array}
\right) $ & $\left( 
\begin{array}{cc}
0 & -1 \\ 
1 & 0
\end{array}
\right) $ \\ \hline\hline
$\left( 
\begin{array}{cc}
0 & -1 \\ 
1 & 0
\end{array}
\right) $ & $\left( 
\begin{array}{cc}
1 & 0 \\ 
0 & -1
\end{array}
\right) $ & $\left( 
\begin{array}{cc}
0 & 1 \\ 
1 & 0
\end{array}
\right) $ & $\left( 
\begin{array}{cc}
0 & 1 \\ 
1 & 0
\end{array}
\right) $ & $\left( 
\begin{array}{cc}
0 & 1 \\ 
1 & 0
\end{array}
\right) $ \\ \hline\hline
$\left( 
\begin{array}{cc}
0 & -1 \\ 
1 & 0
\end{array}
\right) $ & $\left( 
\begin{array}{cc}
0 & -1 \\ 
1 & 0
\end{array}
\right) $ & $-\left( 
\begin{array}{cc}
1 & 0 \\ 
0 & 1
\end{array}
\right) $ & $-\left( 
\begin{array}{cc}
1 & 0 \\ 
0 & 1
\end{array}
\right) $ & $-\left( 
\begin{array}{cc}
1 & 0 \\ 
0 & 1
\end{array}
\right) $ \\ \hline\hline
$\left( 
\begin{array}{cc}
0 & -1 \\ 
1 & 0
\end{array}
\right) $ & $\left( 
\begin{array}{cc}
0 & 1 \\ 
1 & 0
\end{array}
\right) $ & $-\left( 
\begin{array}{cc}
1 & 0 \\ 
0 & -1
\end{array}
\right) $ & $-\left( 
\begin{array}{cc}
1 & 0 \\ 
0 & -1
\end{array}
\right) $ & $-\left( 
\begin{array}{cc}
1 & 0 \\ 
0 & -1
\end{array}
\right) $ \\ \hline\hline
\end{tabular}

\medskip

\begin{tabular}{||l||l||l||l||l||}
\hline\hline
$A$ 4 & $B$ & $BA$ & $\left( BA\right) ^{t}$ & $U$ \\ \hline\hline
$\left( 
\begin{array}{cc}
1 & 0 \\ 
0 & -1
\end{array}
\right) $ & $\left( 
\begin{array}{cc}
1 & 0 \\ 
0 & 1
\end{array}
\right) $ & $\left( 
\begin{array}{cc}
1 & 0 \\ 
0 & -1
\end{array}
\right) $ & $\left( 
\begin{array}{cc}
1 & 0 \\ 
0 & -1
\end{array}
\right) $ & $\left( 
\begin{array}{cc}
1 & 0 \\ 
0 & -1
\end{array}
\right) $ \\ \hline\hline
$\left( 
\begin{array}{cc}
1 & 0 \\ 
0 & -1
\end{array}
\right) $ & $\left( 
\begin{array}{cc}
0 & 1 \\ 
1 & 0
\end{array}
\right) $ & $-\left( 
\begin{array}{cc}
0 & -1 \\ 
1 & 0
\end{array}
\right) $ & $\left( 
\begin{array}{cc}
0 & -1 \\ 
1 & 0
\end{array}
\right) $ & $-\left( 
\begin{array}{cc}
0 & -1 \\ 
1 & 0
\end{array}
\right) $ \\ \hline\hline
$\left( 
\begin{array}{cc}
1 & 0 \\ 
0 & -1
\end{array}
\right) $ & $\left( 
\begin{array}{cc}
1 & 0 \\ 
0 & -1
\end{array}
\right) $ & $\left( 
\begin{array}{cc}
1 & 0 \\ 
0 & 1
\end{array}
\right) $ & $\left( 
\begin{array}{cc}
1 & 0 \\ 
0 & 1
\end{array}
\right) $ & $\left( 
\begin{array}{cc}
1 & 0 \\ 
0 & 1
\end{array}
\right) $ \\ \hline\hline
$\left( 
\begin{array}{cc}
1 & 0 \\ 
0 & -1
\end{array}
\right) $ & $\left( 
\begin{array}{cc}
0 & -1 \\ 
1 & 0
\end{array}
\right) $ & $-\left( 
\begin{array}{cc}
0 & 1 \\ 
1 & 0
\end{array}
\right) $ & $-\left( 
\begin{array}{cc}
0 & 1 \\ 
1 & 0
\end{array}
\right) $ & $-\left( 
\begin{array}{cc}
0 & 1 \\ 
1 & 0
\end{array}
\right) $ \\ \hline\hline
\end{tabular}

Table (1) all the unitary U for the 2-dimension teleportation by Bell channel

$\ \quad $

\end{document}